\documentclass[aps,prl,twocolumn,groupedaddress,floatfix]{revtex4-1}
\usepackage{color}
\usepackage{tabularx}
\usepackage{epsfig}
\usepackage{amsmath}
\usepackage{graphicx}

\begin{document}

\title{The bimodal Ising spin glass in dimension two : the anomalous
  dimension $\eta$}

\author{P. H.~Lundow} \affiliation{Department of Mathematics and
  Mathematical Statistics, Ume{\aa} University, SE-901 87, Sweden}

\author{I. A.~Campbell} \affiliation{Laboratoire Charles Coulomb
  (L2C), UMR 5221 CNRS-Universit\'e de Montpellier, Montpellier,
  F-France.}

\date{\today}

\begin{abstract}
  Direct measurements of the spin glass correlation function $G(R)$
  for Gaussian and bimodal Ising spin glasses in dimension two have
  been carried out in the temperature region $T \sim 1$. In the
  Gaussian case the data are consistent with the known anomalous
  dimension value $\eta \equiv 0$. For the bimodal spin glass in this
  temperature region $T > T^{*}(L)$, well above the crossover
  $T^{*}(L)$ to the ground state dominated regime, the effective
  exponent $\eta$ is clearly non-zero and the data are consistent with
  the estimate $\eta \sim 0.28(4)$ given by McMillan in 1983 from
  similar measurements. Measurements of the temperature dependence of
  the Binder cumulant $U_{4}(T,L)$ and the normalized correlation
  length $\xi(T,L)/L$ for the two models confirms the conclusion that
  the 2D bimodal model has a non-zero effective $\eta$ both below and
  above $T^{*}(L)$.  The 2D bimodal and Gaussian interaction
  distribution Ising spin glasses are not in the same Universality
  class.
\end{abstract}

\pacs{ 75.50.Lk, 05.50.+q, 64.60.Cn, 75.40.Cx}

\maketitle

\section{Introduction}
The square lattice near neighbor random interaction Ising models are
canonical examples of the Edwards-Anderson Ising Spin Glasses (ISG)
\cite{edwards:75}. Numerous studies of these models have been made
over the years, interest being mainly focussed on two versions of this
2D ISG model : the Gaussian model where the random interaction
distribution takes up the continuous Gaussian form and the ground
state is unique, and the bimodal model where positive or negative
interactions ($\pm J$) are randomly distributed, and where the ground
state is massively degenerate. It is well established that both these
models (and any other 2D ISGs) order only at zero
temperature~\cite{ohzeki:09}. (Temperatures will be quoted in units of
$J$).

For any continuous distribution model including the Gaussian the
anomalous dimension critical exponent is analytically known to be
$\eta \equiv 0$ because the ground state is non-degenerate; accurate
and consistent estimates have been made of the correlation length
critical exponent $\nu = 3.52(2)$ for the Gaussian
model~\cite{hartmann:02, carter:02, hartmann:02a, houdayer:04,
  lundow:16, fernandez:16} and for other continuous distribution
models~\cite{lundow:17}. The values of other critical exponents, in
particular the magnetization exponent $\gamma = (2-\eta)\nu$, follow.

For the discrete interaction bimodal model the ordering process is
more complex. It has been established that for finite lattice size $L$
there can be considered to be two regimes separated by a crossover
temperature $T^{*}(L)$, a "ground state plus gap" regime for $T <
T^{*}(L)$, and an "effectively continuous energy level" regime for $T
> T^{*}(L)$~\cite{jorg:06}. In the infinite-$L$ thermodynamic limit
(ThL) $T^{*}$ reaches zero because $T^{*}(L)$ drops with increasing
$L$ as $T^{*}(L) \sim 1.1(1)L^{-1/2}$~\cite{thomas:11,lundow:16}.  In
the strict ground-state limit $T \equiv 0$ and for finite $L$, the
bimodal 2D ISG exponent $\eta$ has been estimated to be $\eta =
0.210(23)$~\cite{ozeki:90} from transfer-matrix ground-state
measurements, $\eta = 0.14(1)$ from ground-state spin
correlations~\cite{poulter:05} and $\eta= 0.22(1)$ from
non-zero-energy droplet probabilities~\cite{hartmann:08}.

There have been consistent estimates over decades indicating a bimodal
ISG effective $\eta \sim 0.20$ in the $T > T^{*}(L)$ regime
also. Initial estimates were from direct finite-temperature
correlation function measurements: first $\eta =
0.4(1)$~\cite{morgenstern:80}, and then from more precise data $\eta =
0.28(4)$~\cite{mcmillan:83}. The latter estimate was qualitatively
confirmed by a Monte Carlo renormalization-group measurement which
indicated $\eta \sim 0.20$~\cite{wang:88} again in what can now be
recognized as being the $T > T^{*}(L)$ regime~\cite{note1}.  Later
numerical simulation estimates were $\eta \sim
0.20$~\cite{houdayer:01}, $\eta \sim 0.138$~\cite{katzgraber:05},
$\eta > 0.20$~\cite{katzgraber:07}, $\eta = 0.20(2)$~\cite{lundow:16}.

These results strongly suggest an anomalous dimension critical
exponent $\eta \sim 0.20$ in both regimes, above and below
$T^{*}(L)$~\cite{note2}, indicating that the bimodal model in the
effectively continuous energy distribution regime $T > T^{*}(L)$ does
not have the same effective exponent $\eta$ as the Gaussian model.
Recent estimates for the bimodal correlation function exponent are
$\nu \sim 5.5$~\cite{thomas:11} and $\nu = 4.8(3)$~\cite{lundow:16},
both significantly higher than the accepted estimate for the Gaussian
model. "Quotient" analyses for the bimodal model~\cite{fernandez:16}
are consistent with the higher value~\cite{lundow:17}.

Nevertheless, the claim has repeatedly been made that the bimodal
model in the $T >T^{*}(L)$ regime is in the same universality class as
the Gaussian model, J\"org {\it et al.}~\cite{jorg:06}, Parisen Toldin
{\it et al.}~\cite{parisen:10,parisen:11}, and Fernandez {\it et
  al.}~\cite{fernandez:16}. It is thus claimed that in the effectively
continuous energy distribution regime $T > T^{*}(L)$ the bimodal
anomalous dimension exponent is $\eta = 0$ and the bimodal correlation
length exponent is $\nu = 3.52(2)$.  The texts of the articles
claiming this Universality do not mention the numerous published
measurements showing a non-zero bimodal model $\eta$ value in the $T >
T^{*}(L)$ regime.

\section{Correlation function measurements}
In 1983 McMillan~\cite{mcmillan:83} carried out direct numerical
measurements of the ISG correlation function
\begin{equation}
  G(R) = \frac{1}{2N} \sum_{i,j}\langle S_{i}S_{j}\rangle^2 \delta(R_{i,j}-R)
\end{equation}
for the 2D bimodal ISG at size $L=96$ as a function of separation $R$
up to $R=16$, for temperatures down to $T=0.89$. As the correlation
function is written $G(R,T) \sim \exp[-R/\xi(T)]/R^{\eta}$ this is a
fundamental defining measurement for $\eta$. It gives a basic
qualitative criterion from which to judge if $\eta$ is zero or not. If
$\eta = 0$, then each plot of $\ln[G(R,T)]$ against $R$ should be a
straight line with a slope proportional to $1/\xi(T)$. If $\eta$ is
not zero, then the $\ln[G(R,T)]$ against $R$ plots should curve
upwards. By inspection, the lowest temperature plots (where the
$\xi(T)$ values are highest) in Ref.~\cite{mcmillan:83} Fig.~1 can be
seen to curve upwards, demonstrating that $\eta$ in the 2D bimodal ISG
is not zero. In more detail, fits to the data provide numerical
estimates for $\eta$ and for correlation lengths $\xi(T)$. McMillan
estimated $\eta = 0.28(4)$ and obtained correlation length estimates
over a narrow temperature range around $T=1$ \cite{mcmillan:83}.  As
remarked by McMillan, a large number of Monte Carlo update steps
($10^6$ in his case) after equilibration are required to achieve
stability of the $G(R,T)$ curves.

As far as we are aware this measurement has never been repeated. We
have been able to extend the measurements to rather lower
temperatures, involving many more update steps, thanks to the
improvements in computing technology over the years (McMillan built
his own computer). The number of updates in the present study cannot
be compared directly with Ref.~\cite{mcmillan:83} as the update
procedures were different.  Unfortunately McMillan was killed in an
accident in 1985 and the tabulated data corresponding to his Fig.~1
are lost. Agreement is however excellent between the points read off
in Ref.~\cite{mcmillan:83} Fig.~1 by eye and the present data for the
same bimodal $G(R,L,T)$.

Data were generated for 2D lattices of linear order $L=16$, $24$,
$32$, $48$, $64$, $96$ with periodic boundary conditions. The spin
interactions $J_{ij}$ were chosen from a bimodal distribution ($\pm 1$
with equal probability) and from a Gaussian distribution
$\mathcal{N}(0,1)$ respectively. During the equilibration phase
standard heat-bath updates were combined with the exchange Monte
Carlo~\cite{hukushima:96} and the Houdayer cluster
method~\cite{houdayer:01} on four replicas. We used $75$ temperatures
in geometric progression between $0.50\le\beta\le 1.50$ and also for
$1\le\beta\le 3$. The same temperature set was thus used on all
lattice sizes. Some $500\ 000$ updates (consisting of a heat-bath
update, temperature exchange and cluster flips) were made in the
equilibration phase for $L=96$. The spin systems were deemed
equilibrated when the estimated specific heat value ($\chi$) only
fluctuated in a random fashion between runs. Once equilibrated we used
only heat-bath updates to measure $G(R)$ after every update.

\section{Gaussian interaction ISG}
\begin{figure}
  \includegraphics[width=3.4in]{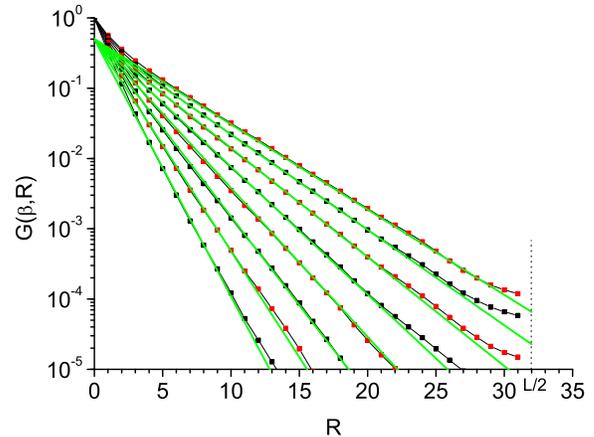}
  \caption{(Color on line) Correlations $G(R)$ as functions of
    distance $R$ for the $L=64$ Gaussian 2D ISG. Inverse temperatures
    $\beta = 0.8$, $0.9$, $1.0$, $1.1$, $1.2$, $1.3$, $1.4$, $1.5$
    from left to right. Green lines : Fits $G(R, \beta) \sim
    \exp[-R/\xi(\beta)]$ } \protect\label{fig1}
\end{figure}

We first show data for the "simple" 2D Gaussian interaction
distribution model where the asymptotic correlation function after a
sufficient number of updates has the zero $\eta$ form $\ln[G(R,
  \beta)] \sim K - R/\xi(\beta)$ with $K \sim -0.7$, at all
temperatures studied. Fig.~\ref{fig1} shows as an example data for
$L=64$ after $10^9$ updates at inverse temperatures from $\beta = 0.6$
to $\beta = 1.5$. It can be seen that the linear asymptotic expression
holds well over the entire range of $R$, except for deviations in the
pre-asymptotic regime at very small $R$ ($R=0$ and $R=1$ essentially),
and when $R$ approaches $L/2$ closely. In the latter limit even after
an infinite number of update steps $G(R)$ must become independent of
$R$ because of well known periodic boundary condition effects :
correlations with "ghost" sites such as $[\pm L,0,0] $ add to the
direct correlation with the site $[0,0,0]$~\cite{talapov:93}. 

\begin{figure}
  \includegraphics[width=3.4in]{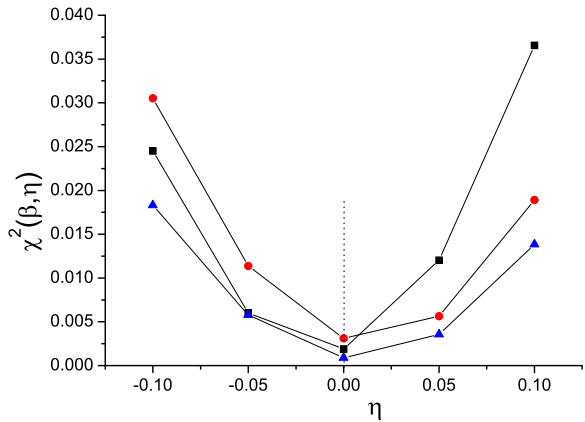}
  \caption{(Color on line) Chi-square fit criterion for fits to the
    $L=64$ Gaussian 2d ISG $G(R)$ data of Fig.~\ref{fig1} at inverse
    temperatures $\beta = 1.5$ (black squares), $1.4$ (red circles),
    $1.3$ (blue triangles). Fits with $\xi(\beta)= 3.57$, $3.3$ and
    $2.78$, and $\eta$ from $-0.10$ to $0.10$.}  \protect\label{fig2}
\end{figure}

In Fig.~\ref{fig2} we apply the chi-square test parameter
$\chi^2=\sum_R (G(R)-G_{\mathrm{fit}}(R))^2/G_{\mathrm{fit}}(R)$ to
fits to the $G(R)$ data, with $R$ from $3$ to $25$ and the three
lowest temperatures of Fig.~\ref{fig1}. With the appropriate
correlation lengths $\xi(\beta)$ fixed and $\eta$ allowed to vary in
the fitting procedure, the chi-square goodness of fit is consistently
at a minimum for $\eta = 0$, the physical value for the Gaussian
model, which is a test of the quality of the measured $G(R)$ data.

\begin{figure}
  \includegraphics[width=3.4in]{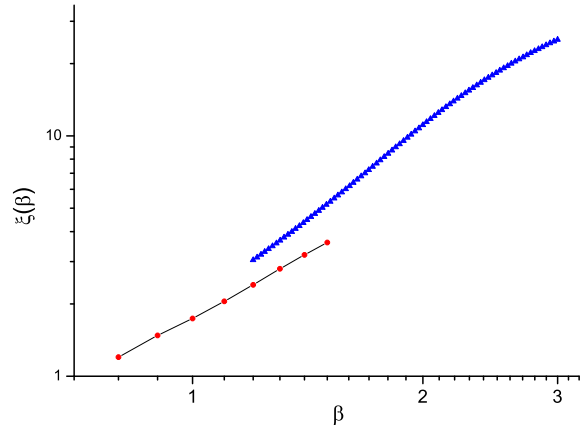}
  \caption{(Color on line)$L=64$ Gaussian 2D ISG. Effective
    correlation lengths as functions of inverse temperature
    $\beta$. Second-moment correlation length from the data generated
    for Ref.~\cite{lundow:16} : blue triangles. "True" correlation
    length, present data, Fig.~\ref{fig1} : red circles.  }
  \protect\label{fig3}
\end{figure}

The estimated correlation length $\xi(\beta)$ increases with
increasing $\beta$ as expected.  The set of measured correlation
lengths $\xi(\beta)$ can be compared to the raw simulation data for
the second-moment correlation length at similar inverse temperatures
which were generated for Ref.~\cite{lundow:16}, Fig.~\ref{fig3}. The
temperature dependence of the two sets is very similar. The offset by
a factor of about $1.15$ can be ascribed to the fact that two
different correlation lengths are being measured; the so-called "true"
correlation length along lattice axes in the present work, and the
second-moment correlation length in Ref.~\cite{lundow:16}. These two
lengths and their differences are discussed for the case of the
canonical 2D and 3D Ising models in Ref.~\cite{butera:04}.  Neither in
the Gaussian case nor in the bimodal case below will we analyse in
terms of an effective correlation length exponent $\nu$, as the range
of temperatures over which measurements were carried out was small and
far from $T=0$ criticality.

\section{Bimodal interaction ISG}
\begin{figure}
  \includegraphics[width=3.4in]{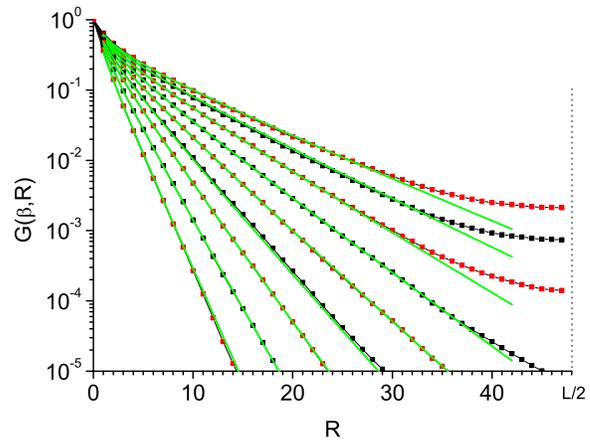}
  \caption{(Color on line) Correlations $G(R)$ as functions of
    distance $R$ for the $L=96$ bimodal 2D ISG. Inverse temperatures
    $\beta = 0.8$, $0.9$, $1.0$, $1.1$, $1.2$, $1.3$, $1.4$, $1.5$
    from left to right. Green curves : fits $G(\beta,R)=
    K\exp[-R/\xi(\beta)]/R^{0.28}$. } \protect\label{fig4}
\end{figure}

\begin{figure}
  \includegraphics[width=3.4in]{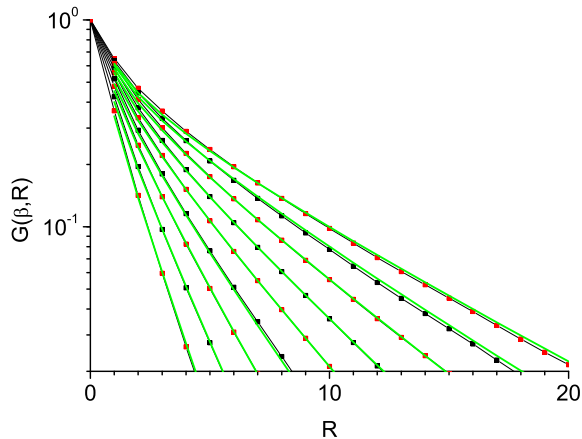}
  \caption{(Color on line) $L=96$ bimodal 2D ISG $G(R)$ small-$R$
    region. Data and fits as in Fig.~\ref{fig4}} \protect\label{fig5}
\end{figure}

The data for $L=96$ after $10^{10}$ updates for inverse temperatures
$\beta = 1.1$ to $1.5$ are shown in Fig.~\ref{fig4}. Other
measurements were made from $\beta = 0.6$ to $\beta = 1.0$ after
$10^9$ updates which was sufficient for these higher temperatures. It
can be seen by inspection of Figs.~\ref{fig4} and \ref{fig5} that
$\ln[G(R, \beta)]$ against $R$ takes the form of curves and not
straight lines, which is a clear qualitative demonstration that for
this model in this temperature range $\eta$ is not zero. It can be
noted that specific heat data Ref.~\cite{thomas:11, lundow:16}
indicate that the "crossover" temperature for $L=96$ is about
$T^{*}(L) \sim 0.12$ so all the present data are well inside the $T >
T^{*}(L)$ "effectively continuous energy level" regime.  

\begin{figure}
  \includegraphics[width=3.4in]{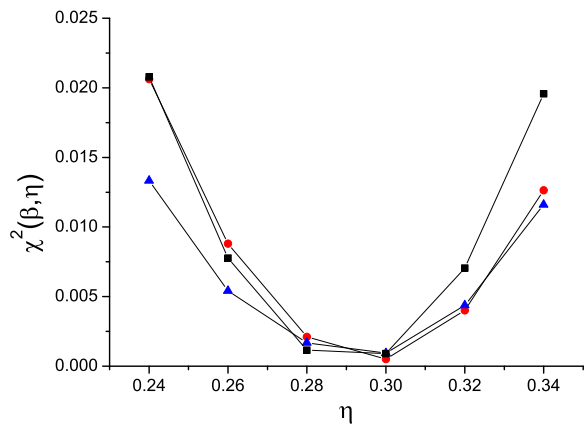}
  \caption{(Color on line) Chi-square fit criterion for fits to the
    $L=96$ bimodal 2d ISG $G(R)$ data of Fig.~\ref{fig5} at inverse
    temperatures $\beta = 1.5$ (black squares), $1.4$ (red circles),
    $1.3$ (blue triangles). Fits with $\xi(\beta)= 7.35$, $6.3$ and
    $5.15$, and $\eta$ from $0.24$ to $0.34$.} \protect\label{fig6}
\end{figure}

In Fig.~\ref{fig6} the same chi-square test procedure as in
Fig.~\ref{fig2} is applied to the bimodal $G(R)$ data for the three
lowest temperatures of Fig.~\ref{fig5}. Here the chi-square goodness
of fit is consistently at a minimum for $\eta$ between $0.28$ and
$0.30$, which is consistent with the McMillan estimate $\eta =
0.28(4)$. A fit with $\eta$ fixed to $0$ can obviously be ruled out.
   
\begin{figure}
  \includegraphics[width=3.4in]{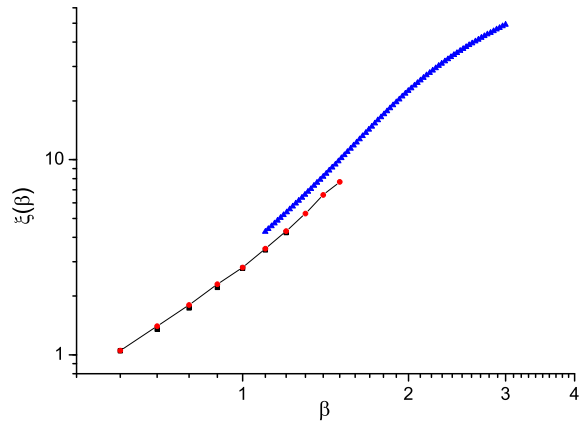}
  \caption{(Color on line) $L=96$ bimodal 2D ISG. Effective
    correlation lengths as functions of inverse temperature
    $\beta$. Second-moment correlation length from the data generated
    for Ref.~\cite{lundow:16} : blue triangles. "True" correlation
    length, present data, Fig.~\ref{fig1} : red circles.}
  \protect\label{fig7}
\end{figure}

Good fits to all the curves have been obtained using a similar
assumption as that made by McMillan, an effective $\eta \sim 0.28$ in
this temperature range.  In Fig.~\ref{fig7} the effective correlation
lengths $\xi(\beta)$ estimated from the fits are compared to the
values obtained from the explicit temperature dependence expression,
Ref.~\cite{mcmillan:83} Eq.~(5), and to the raw bimodal second-moment
correlation lengths from data generated for Ref.~\cite{lundow:16}. The
agreement with McMillan's empirical Eq.~(5) for the true correlation
length (from 35 years ago) is almost perfect. The true correlation
lengths $\xi_{\mathrm{true}}(\beta)$ from the present work and the
second-moment correlation lengths $\xi(\beta)$ for $L=96$ from data
generated for Ref.~\cite{lundow:16} have a qualitatively similar
behavior with an offset such that $\xi_{2m}(T) \sim
1.15\,\xi_{\mathrm{true}}(T)$, just as for the equivalent lengths in
the Gaussian model Fig.~\ref{fig3}.


\section{End points}
As discussed in Ref.~\cite{lundow:17}, for any model (including both
Ising and ISG models) the values at criticality of dimensionless
parameters such as the Binder cumulant $U_{4}(T,L)$ (in the ISG case
$U_{4}=\lbrack\langle q^{4}\rangle\rbrack/\lbrack\langle
q^{2}\rangle\rbrack^{2}$ with $q$ the ISG order parameter defined as
usual by $q = (1/L^{d})\sum_{i} S_{i}^{A}S_{i}^{B}$ where A and B
indicate two copies of the same system and the sum is over all sites)
and the second-moment correlation length ratio $\xi(T,L)/L$ (see
Refs.~\cite{butera:04,katzgraber:05}) depend only on the critical
exponent $\eta$ (see Ref.~\cite{salas:00}).

For any $[T_{c}=0, \eta \equiv 0]$ model such as the 2D Gaussian ISG,
$U_{4}(T,L)$ as a function $\xi(T,L)/L$ is a universal curve
independent of $L$ (except for very small $L$) with a critical $T = 0$
end-point $[\xi(0,L)/L \equiv \infty, U_{4}(0,L)\equiv 1]$. In
contrast, for any $T_{c}=0$ model with a non-zero $\eta$, the
dimensionless parameters saturate with decreasing temperature so there
is a $T=0$ critical end-point with $U_4(0,L) > 1$ and $\xi(0,L)/L \ll
\infty $. (These end-points can be weakly dependent on $L$, but for
deciding if $\eta$ is zero or not this is irrelevant as the $\eta
\equiv 0$ end-point $[\xi(0,L)/L \equiv \infty, U_{4}(0,L)\equiv 1]$
is $L$ independent.)  Consistent data showing these contrasting 2D
Gaussian $\eta \equiv 0$ and 2D bimodal non-zero $\eta$ behaviors have
been made in numerous publications including Katzgraber and
Lee~\cite{katzgraber:05}, Katzgraber {\it et
  al.}~\cite{katzgraber:07}, Parisen Toldin {\it et
  al.}~\cite{parisen:10,parisen:11} and Lundow and
Campbell~\cite{lundow:17}.  Inspection of the raw $\xi(T,L)$ bimodal
data of Fernandez {\it et al.}~\cite{fernandez:16} also shows the same
saturation effect characteristic of a non-zero $\eta$.

\begin{figure}
  \includegraphics[width=3.4in]{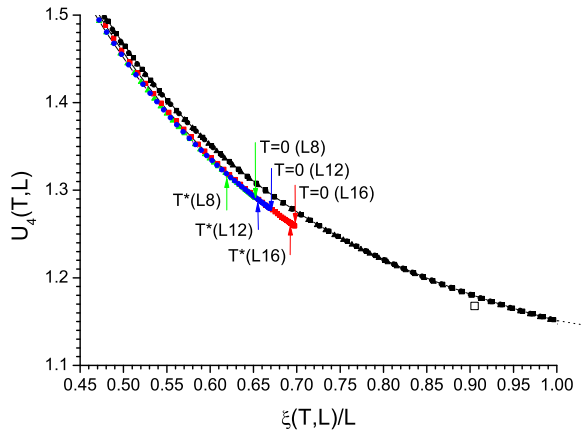}
  \caption{(Color on line) The Binder cumulant $U_{4}(\beta,L)$ as a
    function of the normalized second-moment correlation length
    $\xi(\beta,L)/L$ for the Gaussian model ($L=8$ black triangles,
    $L=12$ black circles, $L=16$ black squares) and for the bimodal
    model ($L=8$ green triangles, $L=12$ blue circles, $L=16$ red
    squares). Zero temperature end-points and crossover temperature
    $T^{*}(L)$ points as indicated. Open square : 2D Ising model
    critical point \cite{salas:00}.}  \protect\label{fig8}
\end{figure}

In Fig.~\ref{fig8} we show $L=8$, $12$ and $16$ Gaussian and bimodal
data to temperatures down to $T = 0.166$. At the higher temperatures
shown the Gaussian and bimodal curves are similar but not identical;
as the temperature drops the Gaussian curve heads towards $[\xi(0,L)/L
  \equiv \infty, U_{4}(0,L)\equiv 1]$ while each bimodal curve comes
to a clear end-point, $[\xi(0,L)/L \sim 0.80, U_{4}(0,L)\sim
  1.30]$. For these sizes the bimodal dimensionless parameters are
essentially saturated by $T=0.166$; the $T=0$ end points and the
crossover temperatures $T^{*}(L)$ (estimated from specific heat
measurements~\cite{lundow:16}) are indicated. The bimodal curves
behave perfectly smoothly through their respective crossover
temperatures $T^{*}(L)$ with no indication whatsoever of a sudden
decrease of the effective $\eta(T)$ around $T^{*}(L)$ as postulated by
J\"org {\it et al.}~\cite{jorg:06}. (On this plot for $L=16$ and
presumably all higher $L$ the $T^{*}(L)$ and $T=0$ points are very
close together.

\section{Discussion}
We can briefly review the articles which suggest that the bimodal
$\eta$ is zero at $T>0$.  J\"{o}rg {\it et al.}~\cite{jorg:06} made
the claim that "The [2D ISG] $T > 0$ properties fall into just one
universality class" on the basis of their numerical scaling plot. It
has already been pointed out in Ref.~\cite{katzgraber:07} that the
Ref.~\cite{jorg:06} Fig.~2 points are derived using an extrapolation
procedure which is not valid for the bimodal model. From a careful
inspection of Ref.~\cite{jorg:06} Fig.~2 it can be seen that the
bimodal (blue) points do not actually lie on the indicated $\eta = 0$
"fit" line. Thus there is no firm basis from the Ref.~\cite{jorg:06}
data for the universality claim.

Parisen Toldin {\it et al.}~\cite{parisen:10,parisen:11} also claim
that "The analysis \ldots confirms that [the Gaussian and bimodal models]
belong to the same universality class". Fig.~1 of
Ref.~\cite{parisen:11} presents data for the Gaussian and bimodal
models, (and for another discrete interaction distribution model) in
the form of plots of the Binder cumulant $U_4(T,L)$ against the
normalized correlation length $\xi(T,L)/L$ as in the present
Fig.~\ref{fig8}. It is stated ``Thus, these [Fig.~1] results [...]
confirm that all models belong to the same universality class." On the
contrary, as discussed in the previous section, our analogous (but
more precise) plots for Gaussian and bimodal data, Fig.~\ref{fig8},
demonstrate that the two models are manifestly in different
universality classes.

Fernandez {\it et al.}~\cite{fernandez:16} also claim that from their
analysis ``Universality among binary and Gaussian couplings is
confirmed to a high numerical accuracy." They explicitly do not
investigate the $T \sim 0$ region nor the crossover" but the same
bimodal end-point behavior as in the present Fig.~\ref{fig8} or
Ref.~\cite{parisen:11} Fig.~1 can be inferred by inspection of the raw
$\xi(L,T)/L$ data of Ref.~\cite{fernandez:16} Fig.~1, leading again to
the same conclusion as in the previous section. The final "high
numerical accuracy" bimodal exponent estimate $|\eta|< 0.02$ of
Ref.~\cite{fernandez:16} (which is incompatible with their "Quotient"
data, reproduced in Ref.~\cite{lundow:17}) can be traced to an
algebraic error in the derivation of their scaling rule expression C1
(see discussion in Ref.~\cite{lundow:17}).

\section{Conclusion}
The present measurements confirm remarkably well the main conclusion
drawn from the pioneering 1983 work of McMillan~\cite{mcmillan:83},
i.e., an effective exponent $\eta \sim 0.28$ for the bimodal 2D ISG at
temperatures in what is now classified as being well in the $T >
T^{*}(L)$ regime. Equivalent 2D Gaussian $G(R)$ data in the same
temperature range can be fitted satisfactorily assuming an effective
exponent equal to the known Gaussian critical value,$\eta \equiv 0$.

A careful comparison of of Binder cumulant $U_{4}(\beta,L)$ against
normalized correlation length $\xi(\beta,L)/L$ Gaussian and bimodal
data show bimodal end-point behavior incompatible with $\eta=0$ so
confirming that the Gaussian and bimodal 2D ISG models are not in the
same universality class \cite{note2}.

Universality in 2D ISGs in the $T > T^{*}(L)$ regime as claimed by
J\"{o}rg {\it et al.}~\cite{jorg:06}, by Parisen Toldin {\it et
  al.}~\cite{parisen:10}, and by Fernandez {\it et
  al.}~\cite{fernandez:16} is incompatible with measurements made by
other groups over 35 years for $T$ both above and below $T^{*}(L)$:
Refs.~\cite{mcmillan:83, wang:88, houdayer:01, katzgraber:05,
  katzgraber:07, lundow:16,lundow:17}, and the present results.

Systems showing dependence of critical exponents on model parameters
are few and far between. Even Baxter's 8-vertex model~\cite{baxter:71}
shows "weak universality", with constant $\eta$, which is not the case
for the 2D ISGs.  A further important implication of the empirical
demonstration of non-universality for ISGs in dimension two is that
there seems no firm reason to suppose that standard universality rules
hold for ISGs in higher dimensions either. Indeed we have presented
strong evidence for breakdown of Universality in ISGs in dimensions
four and five also~\cite{lundow:15,lundow:15a,lundow:17a}.

It was stated fifteen years ago that "classical tools of RGT analysis
are not suitable for spin glasses"~\cite{parisi:01} but this
approach~\cite{castellana:11,angelini:13} has not yet converged on
predictions concerning non-universality.




\begin{thebibliography}{99}
\bibitem{note1} This sophisticated technique \cite{wang:88} appears to
  be very reliable; in the same publication it was validated by 2D
  Ising model measurements, and in addition to the bimodal 2D ISG
  $\eta$ estimate it also provided estimates for bimodal ISG $\eta$ values in
  3D and in 4D which are close to recent estimation
  (thirty years later) from standard simulation measurements.
\bibitem{note2} In the much simpler fully frustrated 2D Ising model
  the ThL \lq\lq effectively continuous energy level\rq\rq{} $T > T^{*}(L)$
  regime effective exponent $\eta$ and the $T \equiv 0$ exponent
  $\eta$ are both equal to $0.5$ \cite{lundow:17}. In 2D the bimodal
  ISG also the $T^{*}(L)$ "transition" does not seem to modify $\eta$.
\bibitem{note3} We find that all continuous distribution 2D
  ISG models constitute a single universality class but that each
  discrete distribution 2D ISG model belongs to an individual class \cite{lundow:17}).
\end{thebibliography}
\end{document}